%% file: conference.tex
\documentclass[conference]{IEEEtran}
\usepackage{cite}
\usepackage{amsmath,amssymb,amsfonts}
\usepackage{algorithm}
\usepackage[noend]{algorithmic}
\usepackage{graphicx}
\usepackage{textcomp}
\usepackage{xcolor}
\def\BibTeX{{\rm B\kern-.05em{\sc i\kern-.025em b}\kern-.08em
    T\kern-.1667em\lower.7ex\hbox{E}\kern-.125emX}}

\makeatletter
\def\ps@IEEEtitlepagestyle{%
  \def\@oddhead{\hbox{}\hfil 2024 International Conference on Hardware/Software Codesign and System Synthesis (CODES+ISSS)\hfil\hbox{}}%
  \def\@evenhead{\@oddhead}%
  \def\@oddfoot{%
    \hbox to \textwidth{%
      \parbox[b]{\columnwidth}{%
        \raggedright\footnotesize
        2832-6474/24/\$31.00~\copyright~2024 IEEE\\
        DOI 10.1109/CODES-ISSS60120.2024.00017
      }%
      \hfil
      \parbox[b]{\columnwidth}{\raggedleft\footnotesize}%
    }%
  }%
  \def\@evenfoot{\@oddfoot}%
}
\makeatother

\begin{document}

\title{\fontsize{20pt}{16pt}\selectfont Special Session: Sustainable Deployment of Deep Neural Networks on Non-Volatile Compute-in-Memory Accelerators}

% \author{\IEEEauthorblockN{1\textsuperscript{st} Yifan Qin}
% \IEEEauthorblockA{\textit{University of Notre Dame} \\
% Notre Dame, US \\
% yqin3@nd.edu}
% \and
% \IEEEauthorblockN{2\textsuperscript{nd} Zheyu Yan}
% \IEEEauthorblockA{\textit{University of Notre Dame} \\
% Notre Dame, US \\
% zyan2@nd.edu}
% % \and
% \linebreakand
% \IEEEauthorblockN{3\textsuperscript{rd} Wujie Wen}
% \IEEEauthorblockA{\textit{North Carolina State Univeristy} \\
% Raleigh, US \\
% wwen2@ncsu.edu}
% \and
% \IEEEauthorblockN{4\textsuperscript{th} Xiaobo Sharon Hu}
% \IEEEauthorblockA{\textit{University of Notre Dame} \\
% Notre Dame, US \\
% shu@nd.edu}
% \and
% \IEEEauthorblockN{5\textsuperscript{th} Yiyu Shi}
% \IEEEauthorblockA{\textit{University of Notre Dame} \\
% Notre Dame, US \\
% yshi4@nd.edu}
% }

\author{
    \textbf{
        Yifan Qin\textsuperscript{$\dagger\circledast$} \ \ \ \ 
        Zheyu Yan\textsuperscript{$\dagger$} \ \ \ \ 
        Wujie Wen\textsuperscript{$\ddagger$} \ \ \ \ 
        Xiaobo Sharon Hu\textsuperscript{$\dagger$} \ \ \ \ 
        Yiyu Shi\textsuperscript{$\dagger*$}
    }\\
    \IEEEauthorblockA{\textsuperscript{$\dagger$}University of Notre Dame, \textsuperscript{$\ddagger$}North Carolina State University \{\textsuperscript{$\circledast$}yqin3, \textsuperscript{$*$}yshi4\}@nd.edu}
}

\maketitle

\begin{abstract}
Non-volatile memory (NVM) based compute-in-memory (CIM) accelerators have emerged as a sustainable solution to significantly boost energy efficiency and minimize latency for Deep Neural Networks (DNNs) inference due to their in-situ data processing capabilities. However, the performance of NVCIM accelerators degrades because of the stochastic nature and intrinsic variations of NVM devices. Conventional write-verify operations, which enhance inference accuracy through iterative writing and verification during deployment, are costly in terms of energy and time. Inspired by negative feedback theory, we present a novel negative optimization training mechanism to achieve robust DNN deployment for NVCIM. We develop an Oriented Variational Forward (OVF) training method to implement this mechanism. Experiments show that OVF outperforms existing state-of-the-art techniques with up to a 46.71\% improvement in inference accuracy while reducing epistemic uncertainty. This mechanism reduces the reliance on write-verify operations and thus contributes to the sustainable and practical deployment of NVCIM accelerators, addressing performance degradation while maintaining the benefits of sustainable computing with NVCIM accelerators.
\end{abstract}

\begin{IEEEkeywords}
in-memory computing, sustainable, neural network, accelerators
\vspace{-10pt}
\end{IEEEkeywords}

\input{introduction}
\input{proposed_method}
\input{experiments}
\input{conclusion}
\input{acknowledgment}

\bibliographystyle{IEEEtran}
\bibliography{references}

\end{document}

%% file: introduction.tex
\section{Introduction}

Deep Neural Networks (DNNs) have revolutionized our society, but their acceleration is hindered by the constant need for data movement between memory and processing units, known as the von Neumann bottleneck \cite{chen2016eyeriss}. Non-volatile memory (NVM) based Computing-In-Memory (CIM) DNN accelerators \cite{shafiee2016isaac} offer a potential solution by enabling parallel in-situ data processing, surpassing CMOS-based counterparts in energy efficiency and density \cite{chen2016eyeriss, zhang2023edge}. These accelerators utilize emerging NVM devices such as ferroelectric field-effect transistors (FeFETs) \cite{reis2018computing}, resistive random-access memories (RRAMs) \cite{qin2020design}, magnetoresistive random-access memories (MRAMs) \cite{angizi2019mrima}, and phase-change memories (PCMs) \cite{sun2021pcm}, providing a sustainable solution for DNN inference acceleration. Despite their advantages, NVM devices in NVCIM DNN accelerators suffer from inherent non-idealities, such as device variations \cite{qin2020design, yan2023improving}. These variations cause perturbations in device conductance after programming \cite{rizzi2011role}, often resulting in Gaussian-distributed conductance values \cite{qin2020design}. Consequently, the model weights are affected, ultimately impacting the inference accuracy of NVCIM DNN accelerators \cite{qin2020design, yan2021uncertainty, yan2023improving}.

\begin{figure}[t]
  \centerline{\includegraphics[scale=0.27]{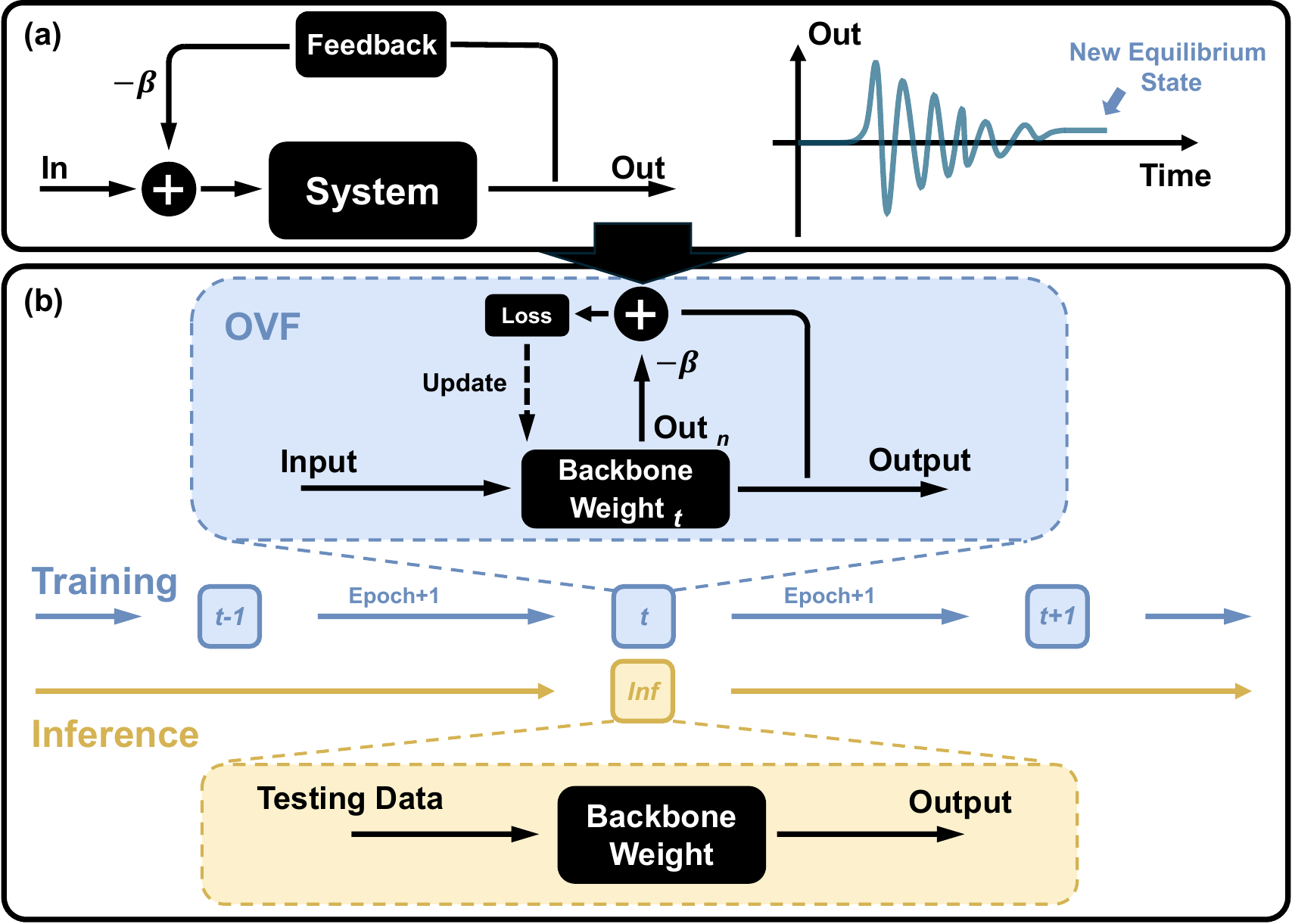}}
  \caption{(a) Schematic diagram of a classic negative feedback system and the process of the system setting to a new equilibrium state. (b) Illustration of negative optimization training mechanism.}
  \label{fig: overview}
  \vspace{-15pt}
\end{figure}

Ensuring reliable DNN inference on unreliable NVM substrates presents a significant challenge. Among solutions \cite{degraeve2015causes, shim2020two, eldebiky2023correctnet}, hardware write-verify has emerged as a widely adopted method for accelerator deployment. However, the time-consuming and energy-intensive iterative write and verify operations hinder sustainable deployment. Therefore, we need more robust network models to achieve more sustainable deployment and acceleration.

Noise-injection training \cite{yan2023improving} is widely used to enhance model robustness by exposing DNNs to Gaussian noise during training. This improves the model's noise tolerance. However, state-of-the-art (SOTA) noise-injection methods have limitations, such as limited accuracy improvement, increased epistemic uncertainty \cite{gawlikowski2023survey}, and challenges in achieving convergence. We attribute this to the mismatch between the non-deterministic nature of noise and the deterministic nature of training. Specifically, \textbf{first}, the network is exposed to only a finite number of noise samples during training, limiting its ability to fully understand noise patterns. \textbf{Second}, random noise samples provide diverse optimization directions, but some may lead to incorrect states, increasing uncertainty and hindering convergence.

We believe this mismatch can be mitigated by acquiring sufficient variation information during training, rather than relying solely on the final output, as is common in SOTA methods. This hypothesis is rooted in modern control theory, where stability relies on negative feedback. When a system is subjected to noise, noisy outputs help the system resist perturbation through negative feedback, achieving a new equilibrium state. Figure~\ref{fig: overview}(a) illustrates this process. Feedback is generated from a portion of the outputs and modulated by the negative feedback coefficient $\beta$. Inspired by this, we introduce a novel \textbf{negative optimization training mechanism}, which incorporates negative contributions from outputs, reduces the impact of noise, and helps the neural network achieve a more robust state, as shown in Figure~\ref{fig: overview}(b). At a high level, the entire negative optimization training mechanism can be summarized as follows, we use negative optimization training to enhance the robustness of the DNN backbone. After training, all negative contribution components are removed, leaving a robust and unaltered DNN backbone model.

Here we provide an implementation called the \textbf{Oriented Variational Forward (OVF)} training method to ground this mechanism. During training, variational inference outputs $Out_n$ are generated with the same backbone weights and oriented increasing amplitude noise, and contribute negatively to the objective in back-propagation with coefficient $\beta$, thereby enhancing DNN robustness. The ``negative'' aspect of OVF reduces the bad effects of target variation while preserving its subject status, and the ``feedback'' component introduces supplementary noise information that differs from what the backbone output contains, contributing to the objective. OVF constrains network optimization based on the influence of noise itself, with stronger constraints corresponding to greater perturbations. This ensures the network does not deviate far from the optimal direction, facilitating stable convergence to the optimal state during the training process.

Our contributions can be summarized as follows:
\begin{itemize}
    \item We introduce a negative optimization training mechanism into the DNN training process to enhance stability and improve robustness to device variations. \textit{To the best of our knowledge, this is the first approach of its kind.}
    \item We propose a novel implementation of this mechanism: Oriented Variational Forward (OVF), which optimizes the network from a comprehensive variational performance perspective.
    \item Our simulations on NVCIM DNN accelerators with varying device variations demonstrate the effectiveness of OVF in mitigating sensitivity and output fluctuations. OVF boosts confidence and convergence probability while reducing epistemic uncertainty. For example, it achieves up to a 46.71\% improvement in DNN average inference performance compared to state-of-the-art methods.
\end{itemize}

%% file: proposed_method.tex
\section{Proposed methodology}

In this section, we introduce the negative optimization with the OVF training method.

Our proposed method is inspired by negative feedback theory, a cornerstone of system control. We treat weight variations as perturbations and aim to improve system robustness by suppressing this ``noise'' through negative optimization. The primary challenge is constructing an effective negative constraint. Simply employing a negatively scaled output of the DNN, as in standard negative feedback systems, is insufficient because it only scales the loss function without altering the training method.

Instead, we require negative constraints that can track changes in the output while remaining distinct from it. The constraint must satisfy two criteria: \textbf{First}, it should be generated by components influenced by the same noise pattern present in the backbone. \textbf{Second}, the negative constraint should strongly connect with the backbone weights, accurately reflecting weight perturbations.

\begin{figure}[t]
  \centerline{\includegraphics[scale=0.3]{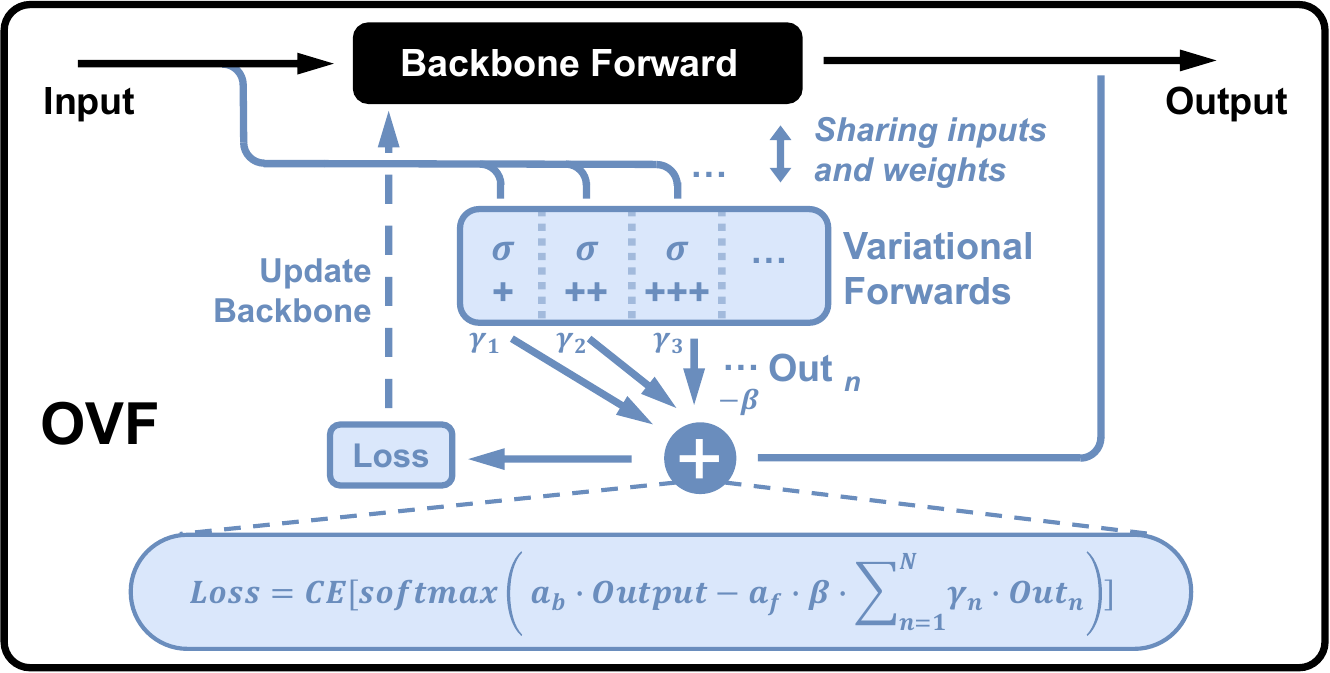}}
  \caption{Negative optimization implementations: oriented variational forward training.}
  \label{fig: ovf_method}
  \vspace{-15pt}
\end{figure}

As shown in Figure~\ref{fig: ovf_method}, OVF generates constraints using less representative outputs $Out_{n}$ from oriented variational forwards, which involve device variations larger than those in backbone inference. By employing negative constraints, OVF prevents the backbone from deviating from the optimal optimization direction. Specifically, during each training iteration, we sample a variation instance $\Delta\mathbf{w}_i$ from a Gaussian distribution $\mathcal{D}ist=\mathcal{N}(0,\sigma^2)$, which is the same as the inference device variations in accelerators. This variation is added to the backbone weights in the feed-forward process, generating the backbone output $\mathcal{O}_\text{backbone}$. Unlike typical training, which directly performs back-propagation and weight update, OVF performs multiple oriented variational forwards using the same variation-free backbone weights with noises sampled from $\mathcal{N}(0,\sigma^2)$ but with larger $\sigma$, collecting $N$ constraint outputs $Out_n$. 
% To achieve accurate predictions, the constraint deviating further from the target should have a more significant negative impact during training, corresponding to decay factors $\gamma_n$.
The total output $\mathcal{O}_\text{total}$ is given by
\begin{equation}
\label{total_out}
\mathcal{O}_\text{total}=a_{b}\cdot\mathcal{O}_\text{backbone}-a_{f}\cdot\beta\cdot\sum_{n=1}^{N}\gamma_n\cdot Out_{n}
\end{equation}
where $\beta$ is the negative constraint coefficient, $\gamma_n$ are the decay factors, and $a_{b}$ and $a_{f}$ are factors influencing the contribution of the two-part outputs. The back-propagation process commences only after all outputs have been obtained. 
% Further details regarding this procedure can be found in Algorithm~\ref{OVF_algo}.

When selecting hyperparameters, it is important to note that as the standard deviation $\sigma$ of the Gaussian distribution increases, noise instances introduce a higher degree of entropy and uncertainty to the model and its outputs. This leads to more significant deviations from the target. Consequently, $Out_n$ generated with a larger $\sigma$ is assigned a larger decay factor $\gamma_n$ in Eq.~\ref{total_out}, thereby imposing a stronger constraint on the backbone model. To achieve this, we set $\gamma_n$ as $10^{n-N}$. These variational forwards generate constraints from the same variation-free backbone weight and employ the same Gaussian noise pattern as the backbone variational forward with different parameters, thus satisfying the criteria outlined previously.

%% file: experiments.tex
\section{experiments}

In this section, we introduce the weight variation model and the setup of experiments, then show the effectiveness of the OVF method with experimental results.

\subsection{Model and Setup}\label{sec: model}

Without loss of generality, we primarily consider device variations stemming from the programming process, where the programmed conductance value in NVM devices deviates from the desired value.
Set a DNN weight with $M$ bits, the desired weight value $\bar{\mathcal{W}_d}$ after quantization can be expressed as
\begin{equation}
\bar{\mathcal{W}_d} = \frac{\max{|\mathcal{W}|}}{2^M - 1}\sum_{i=0}^{M-1}{m_i \times 2^i}
\end{equation}
where $\mathcal{W}$ represents floating-point weights, $\max{|\mathcal{W}|}$ denotes the maximum absolute value among weights, and $m_{i} \in \{0,1\}$ signifies the value of the $i^{th}$ bit of the desired weight value. For a NVM device representing $K$ bits of data, each weight can be stored in $M/K$ devices\footnote{For simplicity, we assume that $M$ is a multiple of $K$.}, and the mapping process is given by $\bar{g_{j}} = \sum_{i=0}^{K -1} m_{j\times K+i} \times 2^i$, where $\bar{g_{j}}$ is the desired conductance of the $j^{th}$ device. It is worth noting that negative weights can be mapped in the same manner to a separate crossbar array. Taking device variation into account, the actual device conductance after programming is denoted as $g_{j} = \bar{g_{j}} + \Delta{g}$, where $\Delta{g}$ represents the deviation from the desired conductance value $\bar{g_j}$ and follows a Gaussian distribution. Consequently, the actual weight $\mathcal{W}_{p}$ represented by programmed NVM devices is given by
\begin{equation}
\mathcal{W}_{p} =\bar{\mathcal{W}_d} + \frac{\max{|\mathcal{W}|}}{2^M - 1} \sum_{j=0}^{M/K-1} {\Delta{g} \times 2^{j\times K}}\
\end{equation}

In our study, we set $K=2$, while the value of $M$ was determined by the specific model configuration. For this research, we selected $M=8$, indicating 8-bit precision for a single DNN weight and 2-bit precision for a single device conductance. To model device variation, we employed a Gaussian distribution with $\Delta{g} \sim \mathcal{N}(0, \sigma_{d}^{2})$, where $\sigma_d$ represents the relative standard deviation of conductance corresponding to the maximal conductance of a single device. We set a constraint on $\sigma_d$, limiting it to $\sigma_d \leq 0.4$. This range is considered reasonable in prior research \cite{reis2018computing, qin2020design, angizi2019mrima, sun2021pcm} and can be achieved through optimizations at the device level, including write-verify techniques.

We carried out experiments using the PyTorch environment on NVIDIA GPUs. Unless otherwise specified, the reported results represent the average of at least five independent runs. We used the average accuracy of noise-injection inference as the performance metric and performed a Monte Carlo simulation with 200 runs to ensure high precision. Our experiments show that the results have a 95\% confidence interval of $\pm 0.01$, in line with the central limit theorem. We compared OVF with two baselines: 1) vanilla training (W/O Noise) and 2) Gaussian noise-injection training (W/ Noise). We did not evaluate OVF against other orthogonal methods, such as NAS-based DNN topology design or Bayesian Neural Networks, as it can be used in combination with them.

Through our experiments with comprehensive datasets and neural network backbones, we have found that the appropriate value of the negative constraint coefficient $\beta$ consistently falls within the set $\{1e-1, 1e-2, 1e-3, 1e-4\}$. Consequently, a four-step search suffices to determine the setting. For hyperparameter values, we set $start = 0$ and $end = 2 \times \sigma_d$, evaluating OVF efficacy across various $\sigma_d$ values. We also set the contribution factors $a_b = a_f = 1/(N+1)$, where $N$ represents the number of variational forwards. Other training hyperparameters, such as learning rate, batch size, and learning rate schedulers, follow best practices for training a noise-free model.

\subsection{Accuracy Improvement}
% \vspace{-10pt}

\begin{figure}[h]
  \centerline{\includegraphics[scale=0.39]{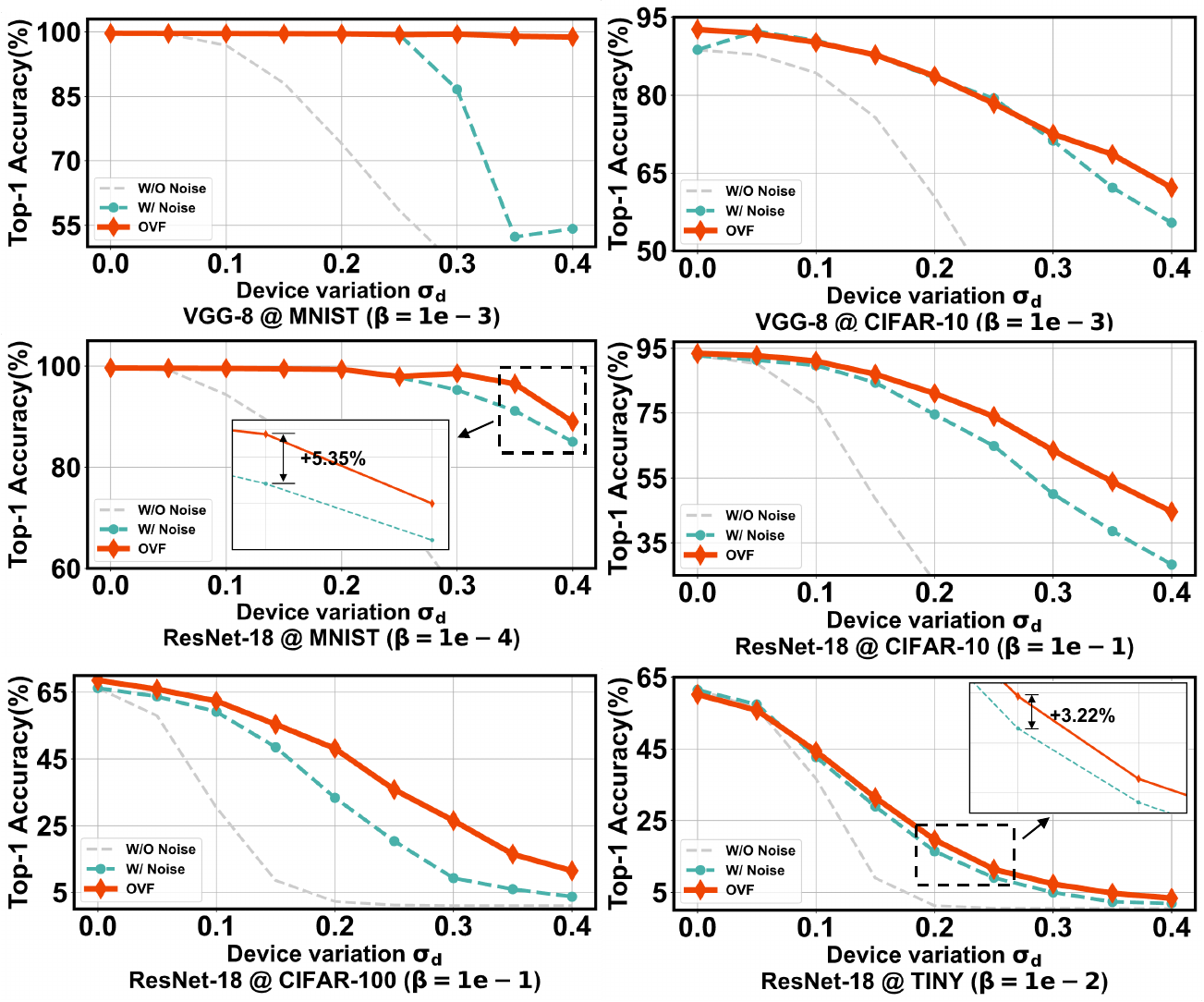}}
  \caption{Effectiveness of OVF: Average noisy inference accuracy on VGG-8 and ResNet-18 backbone models for different datasets across $\sigma_d$ values.}
  \label{fig: OVF_results}
  \vspace{-5pt}
\end{figure}

For our experiments, we employed the VGG-8 backbone and the ResNet-18 backbone on the MNIST, CIFAR-10, CIFAR-100, and Tiny ImageNet datasets. In OVF, we empirically set the variational forwards $N=3$ for both VGG-8 and ResNet-18, with each increase $\Delta\sigma_d$ fixed at 0.05. Furthermore, the negative constraint coefficient $\beta$ for each model on a specific dataset was determined through a four-step search process, as stated in section \ref{sec: model}.

Figure~\ref{fig: OVF_results} illustrates the Top-1 inference accuracy of models trained with different methods under varying levels of device variations $\sigma_d$, following the noise model discussed in Section~\ref{sec: model}. OVF clearly surpasses all baselines in most device value deviation values and performs similarly to baselines in rare cases where the device variation is too small to have a significant impact. Compared to the Gaussian noise-injection training baseline, OVF enhances the Top-1 accuracy by up to 46.71\%, 6.78\%, 5.35\%, 16.30\%, 17.21\%, and 3.22\% in VGG-8 for MNIST and CIFAR-10, ResNet-18 for MNIST, CIFAR-10, CIFAR-100, and Tiny ImageNet, respectively. OVF constrains the network based on the overall forward performance and is particularly suitable for networks that have not reached the limit of their representational ability, such as VGG-8 on MNIST.

The effectiveness of the OVF method highlights the generality and practicality of the negative optimization training mechanism in enhancing DNN robustness against device variation, thereby contributing to the sustainable deployment of NVCIM accelerators.

\subsection{Uncertainty and Convergence}

\begin{figure}[h]
  \centerline{\includegraphics[scale=0.25]{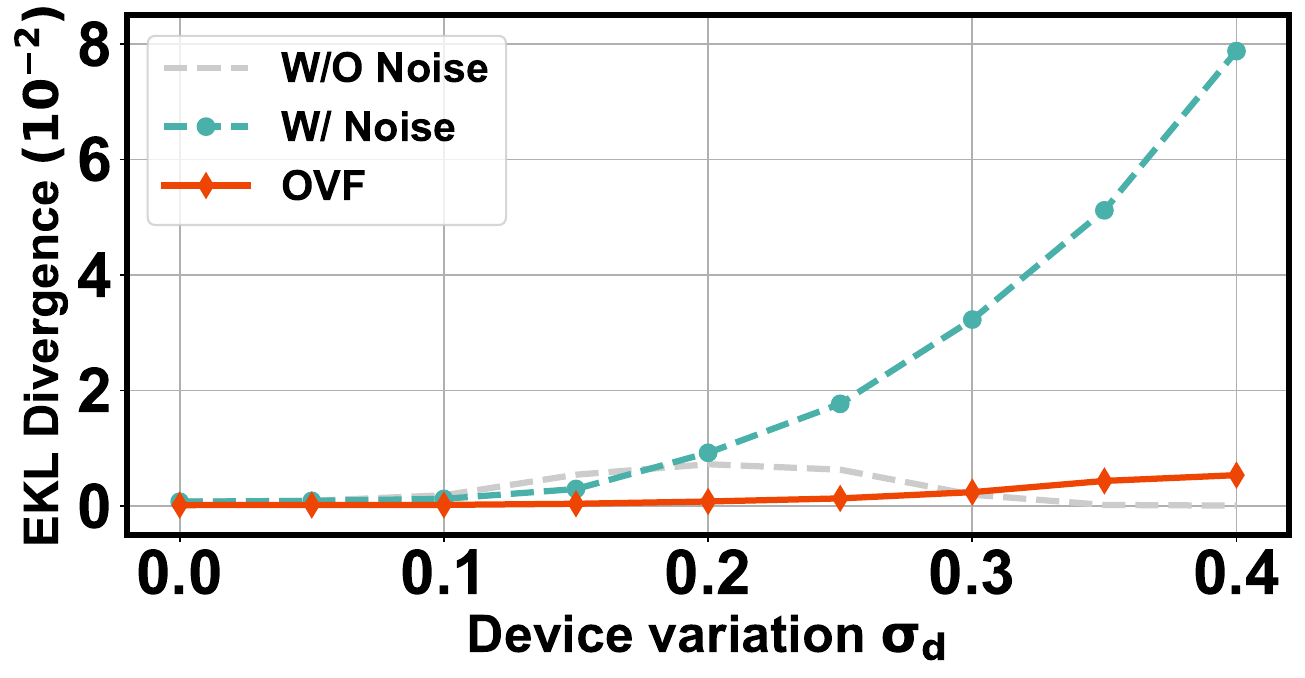}}
  \caption{Average EKL divergence for correct predictions with different methods.}
  \label{fig: EKL_results}
  % \vspace{-10pt}
\end{figure}

Device variation amplifies epistemic uncertainty, resulting in increased output uncertainty. To quantify the impact of device variation on uncertainty, we employ the Expected Kullback–Leibler (EKL) divergence \cite{gawlikowski2023survey} (lower values indicate better performance). Accuracy is excluded from the analysis for fair comparison purposes. Specifically, among all correct predictions in noisy inference, we calculate the Kullback–Leibler divergence between each softmax output and its corresponding label. The results, shown in Figure~\ref{fig: EKL_results}, represent the average EKL divergence for each correct prediction. Compared to the W/O Noise baseline, the W/ Noise baseline improves accuracy but at the cost of increased uncertainty. In contrast, our OVF method not only achieves even higher accuracy than the noise-injection training baseline but also maintains low uncertainty and high confidence in the output. In cases where device variation is too substantial for effective predictions using the vanilla training baseline, its EKL divergence appears slightly lower than that of OVF, as it generates completely random and meaningless predictions.

For certain devices and aging-related issues, the device variation can be significant. The low uncertainty achieved by OVF also contributes to model convergence. For example, in 10 separate runs of experiments with VGG-8 on MNIST using $\sigma_d=0.35$, the number of non-converging models\footnote{where accuracy decreases by more than 5\% compared to the average accuracy across multiple independent runs} is 6 and 0 for noise-injection training and OVF, respectively. In addition to the increase in accuracy, this may also partially explain the substantial improvement of OVF in VGG-8 for MNIST, as shown in Figure~\ref{fig: OVF_results}.

%% file: conclusion.tex
\section{conclusion}

In conclusion, the proposed Oriented Variational Forward (OVF) method significantly enhances the robustness of deep neural networks (DNNs) against device variations and contributes to the sustainable deployment of NVCIM accelerators. By maintaining high accuracy while keeping uncertainty low, OVF reduces the reliance on write-verify operations, improving deployment time and energy efficiency, and achieving better inference accuracy on chips with the same devices. This method demonstrates the generality and practicality of the negative optimization training mechanism, providing valuable insights into the development of robust AI accelerators that can adapt to the non-ideal characteristics of NVM devices, thereby promoting the sustainable development of AI hardware.

%% file: acknowledgment.tex
\section*{Acknowledgment}

This research was supported by ACCESS – AI Chip Center for Emerging Smart Systems, sponsored by InnoHK funding, Hong Kong SAR.